\documentclass{svproc}
\usepackage{url}

\usepackage{graphicx}

\usepackage{mathrsfs}
\usepackage{amssymb}
\usepackage{amsmath}
 
\usepackage{enumitem} 
\usepackage{tikz} 
\usepackage{graphicx,dsfont}       
\usepackage{subfigure}  

\numberwithin{equation}{section}
\numberwithin{definition}{section}
\numberwithin{theorem}{section}
\numberwithin{remark}{section}
\usepackage{expl3}
\usepackage{xcolor}
\usepackage{lmodern,graphicx,amsmath,mathtools,xfrac,array,multirow,booktabs,hyperref}
\usepackage[only, llbracket, rrbracket]{stmaryrd}  
\usepackage{tcolorbox}

\newcommand{\abs}[1]{\lvert#1\rvert}

\newcommand \diag {\text{diag}}
\newcommand \Kasner {\text{Kasner}}
\newcommand \Vol {\text{Vol}}
\newcommand \Ibf 	{\mathbf I} 
\newcommand \Kmoi {{K^-}{}}
\newcommand \Kpoi {{K^+}{}}
\newcommand \phimoi {\phi^-}
\newcommand \phipoi {\phi^+}
\newcommand \gmoi {{g^-}{}}
\newcommand \gpoi {{g^+}{}}

\newcommand \Tr {\text{Tr}}
\newcommand \Sbf {\mathbf S}
\newcommand \Ucal   {\mathcal U}
\newcommand \bei {\begin{itemize}}
\newcommand \eei {\end{itemize}}
\newcommand \Mcal {\mathcal M}
\newcommand \Lcal {\mathcal L} 
\newcommand \be         {\begin{equation}}
\newcommand \ee         {\end{equation}} 
\newcommand \RR         {\mathbb{R}} 
\newcommand \del        \partial
\newcommand \eps    \varepsilon  
\newcommand \auth {\textsc}  
\newcommand \Hcal {\mathcal H}    
\newcommand \Tbb {\mathbb T} 
\newcommand \Kcirc {\mathring{K}} 

\begin{document}

\mainmatter            

\title{Singularity scattering laws for bouncing cosmologies: 
a brief overview} 

\titlerunning{Singularity scattering laws for bouncing cosmologies}   
%
\author{Philippe G. LeFloch\inst{\star}} 
\authorrunning{Philippe G. LeFloch}  

\tocauthor{Philippe G. LeFloch} 

\institute{Laboratoire Jacques-Louis Lions, 
\\
Centre National de la Recherche Scientifique, 
Sorbonne Universit\'e, 
\\
4 Place Jussieu, 75252 Paris, France.
\\
\email{contact@philippelefloch.org}
\\
\, Blog: \textit{philippelefloch.org}
}

\maketitle     

\begin{abstract} For contracting/expanding bouncing cosmologies, the formulation of junction conditions at a bouncing 
was recently revisited by the author in collaboration with B. Le Floch and G. Veneziano. The regime of interest here is
 the so-called quiescent regime, in which a monotone behavior of the metric is observed and asymptotic expansions can be derived. 
 Here, we overview our new methodology based on the notion of {\sl singularity scattering maps} and {\sl cyclic spacetimes}, 
 and we present our main conclusions. In particular, we provide a classification of all allowed bouncing junction conditions, including {\sl three universal laws}. 
\end{abstract}


\section{Introduction}
\label{section-intro}

\paragraph{Self-gravitating matter fields.}

We overview recent developments \cite{LLV-1a,LLV-2,LLV-1b} on weak solutions to Einstein's field equations, established 
in collaboration with B.~Le~Floch (ENS, Paris) and G. Veneziano (CERN, Geneva), 
concerning the formulation of scattering laws allowing one to pass from a contracting phase to an expanding one, across a singularity hypersurface. The interest of the author for weak solutions to the Einstein equations began fifteen years ago
\cite{PLFMardare,PLFRendall,PLFSormani,PLFStewart}, and continued until recent years in collaboration with B.~Le~Floch
 \cite{LeFlochLeFloch-1}--\cite{LeFlochLeFloch-4}. 
 
 Since this is only a brief overview of the subject, only a few results and comments are discussed and the reader is referred to the cited papers for further background material and results on the subject. 

We are interested in self-gravitating matter fields and physically realistic models, whose solutions may involve (possibly impulsive)
gravitational waves (deformations of the spacetime geometry), shock waves (in the fluid), and phase transition interfaces (for complex fluids). 
Such waves are represented by singularity hypersurfaces across which the solutions to the Einstein equations 
exhibit a jump discontinuity or even a blow-up.  We need various techniques of geometric analysis and mathematical physics, including 
Lorentzian geometry (in presence of metrics with weak regularity) and arguments from
the theory of partial differential equations (of nonlinear hyperbolic or elliptic type). We also pay attention to the underlying
 physical modeling, including (possibly modified) gravity modeling and continuum physics. 

\paragraph{New methodology.}

In  \cite{LLV-1a}, we proposed a new (mathematical) perspective on bouncing cosmologies, that is, spacetimes containing 
contracting and expanding phases of big crunch and big bang type connected together by a bounce, which is 
regarded as a singularity hypersurface across which the small-scale physics has been ``factored out'' (see Section~\ref{section--43} ). 
On this subject, a very large literature exists, for instance
 by Ashtekar, Brandenberger, Ijjas, Gasperini, L\"ubbe, Pawlowski,  Penrose, Peter, Steinhardt, Tod, Turok, Veneziano, Wilson-Ewing, and others. We do not attempt to review this literature  and refer the reader to \cite{Ashtekar,AshtekarWilsonEwing,BV,Brizuela:2009nk,PenroseCCC1,SteinhardtTurok2004,Tod,Wilson-Ewing-LQC} and the references cited therein. 

The framework proposed in \cite{LLV-1a,LLV-2,LLV-1b} relies on a systematic study of bouncing junctions  at geometric and fluid interfaces. For regular junction hypersurfaces, one can use Israel's junction conditions \cite{BarrabesIsrael}, 
about which we refer the reader to Marc and Senovilla \cite{MarsSenovilla}. 

To deal with singularity hypersurfaces, we begin by analyzing the degrees of freedom and constraints. 
In the regime of 
``quiescent'' cosmology (cf.~Barrow~\cite{Barrow} and Andersson and Rendall~\cite{AnderssonRendall}),
 spacetimes have a monotone behavior (as opposed to BKL oscillations identified by Belinsky, Khalatnikov, Lifshitz)
 and asymptotic expansions of Fuchsian type can be established.
The quiescent behavior on gravitational singularities is observed for large classes of matter models as well as for the vacuum Einstein equations in high dimensions, or for spacetimes admitting certain symmetries (for instance $\Tbb^2$ symmetry).

A {\sl classification of bouncing laws} is established in \cite{LLV-1a}, which is based on 
analyzing the scattering phenomena near a singularity hypersurface  and
 formulating junction conditions via universal or model-dependent laws. This also naturally leads to 
 the construction 
 of 
 cyclic spacetimes describing the collision of two gravitational waves beyond singularities, and to
 the  resolution of the {\sl global plane-wave collision problem,} as we call it.  
 
 We will not discuss the plane-symmetric problem here and we refer the reader to \cite{LLV-1b} as well as the earlier works \cite{FloresSanchez:2003,FloresSanchez:2008}. 

 \paragraph{Outline of this paper.}
 
In Section~\ref{section--2}, a brief presentation of recent results in mathematical general relativity is given. 
In Section~\ref{section--3}, we discuss our methodology in order to deal with spacetimes with singularity hypersurfaces. 
Next, in Section~\ref{section--4}, we introduce the notions of scattering maps and cyclic spacetimes and we state 
our local existence theory. 
Finally, Section~\ref{section--5} is devoted to the presentation of the classification of scattering maps. 


\section{Global nonlinear stability of Einstein spacetimes}
\label{section--2} 
 
\subsection{Background}
\label{section--21} 
 
\paragraph{The initial value problem.}
 
The evolution problem for the Einstein equations is formulated as follows. The unknown is a Lorentzian four-manifold $(M,g)$ with signature $(-1,1, 1, 1)$ satisfying the field equations 
\be \label{equa-EE} 
G = 8 \pi \, T, 
\ee
supplemented with a prescribed data set representing the initial geometry and matter content. 
We are given a Riemannian $3$-manifold $(M_0, g_0, k_0)$, representing a
 hypersurface embedded in the spacetime, together with a scalar matter density field $\rho_0$ and a vector field $J_0$, which should satisfy 
 Einstein's constraint equations, namely the Hamiltonian equation
\be
\textbf{Scal}_{g_0} + |k_0|^2 - \textbf{Tr} (k_0^2) = 16 \pi \rho_0 
\ee
and the momentum equation
\be
\textbf{div}_{g_0} \big ( k_0 - \textbf{Tr} (k_0) g_0 \big) =  8 \pi \, J_0.
\ee
Under a suitable gauge choice, the last two equations form a nonlinear elliptic equation while 
from the Einstein equations \eqref{equa-EE}  one can also extract a nonlinear wave system satisfied by the metric. 

\paragraph{The global nonlinear stability problem.} 

Let us review a few results on the nonlinear stability of vacuum spacetimes under small perturbations. The theory was 
restricted to vacuum spacetimes having $T=0$ until recently (see next paragraph)
and has a long history beginning in the 1990  with the pioneering contribution 
on Minkowski spacetime     
by Christodoulou and Klainerman, followed by important work by Bieri, Lindblad, Rodnianski, Hintz, Vasy, and others. 
More recently, further advances were made concerning the nonlinear stability of the Schwarzschild (stationary black hole) and Kerr (rotating)  black hole spacetimes (Dafermos, Holzegel, Klainerman, Szeftel, Rodnianski, and others). In these works, the global dynamics of small perturbations of a given geometry are studied, and the analysis relies on numerous mathematical techniques 
for nonlinear wave equations and nonlinear elliptic equations: linearized stability, dispersive estimates, nonlinear structure, time decay, etc. 
 
\paragraph{Matter fields, low decay, and singularities.}

Our objective in the present overview is to go {\sl beyond vacuum} spacetimes. Several directions are of interest. 
In analyzing matter spacetimes as well as spacetimes beyond asymptotic symmetry. The author recently 
treated three types of interrelated problems:
the nonlinear stability of Klein-Gordon fields \cite{LeFlochMa1}--\cite{LeFlochMa4},
the Einstein constraints beyond spherical symmetry \cite{LeFlochNguyen},
and the evolution in presence of singularity hypersurfaces \cite{LLV-1a,LLV-2,LLV-1b}, which is our main aim for the present overview. 
 

\subsection{Self-gravitating massive matter field}  

\paragraph{The global dynamics of massive fields.}

In presence of a massive scalar field, the Einstein equations exhibit a very complex dynamics, 
and analyzing the decay properties at timelike, null, and spacelike infinity is a very challenging problem. 
The first results on the nonlinear stability of self-gravitating massive matter fields
were established in recent years 
by LeFloch and Ma~\cite{LeFlochMa4}
 and 
by Ionescu and Pausader~\cite{IonescuPausader0,IonescuPausader}. Two independent and very different proofs are thus 
now available.  The method proposed by Ionescu and Pausader 
is based on the technique of spacetime resonances, originally developed (for simpler wave problems) 
by Shatah, Germain, and Masmoudi; see the references in~\cite{IonescuPausader0,IonescuPausader}. 
The simpler class of solutions coinciding with the Schwarzschild spacetime outside a light cone was analyzed earlier in independent works by LeFloch and Ma \cite{LeFlochMa2}--\cite{LeFlochMa3} and by Wang \cite{Wang}. 

In \cite{LeFlochMa1} and then in \cite{LeFlochMa4}, a new vector field method is introduced, which we call the ``Euclidian-hyperboloidal foliation method''
and is relevant in order to solve the global existence problem for a broad class of 
coupled systems of nonlinear wave and Klein-Gordon equations. 
  For a precise statement we refer the reader to \cite{LeFlochMa4} and we only sketch our main conclusion here.  
  
\paragraph{Nonlinear stability of self-gravitating massive fields: informal version}

Consider the Einstein equations coupled to a Klein-Gordon field $\phi$, satisfying therefore the evolution equation
\be
- \Box_g \phi + m^2 \, \phi = 0. 
\ee
Let $(M_0 \simeq \RR^3, g_0, k_0, \phi_0, \phi_1)$ be an initial data set that is assumed to be sufficiently close to 
(vacuum) Minkowski data and 
to enjoy certain (possibly slow) decay conditions at spacelike infinity (for instance, 
possibly non-spherically symmetric at infinity). Suppose also that these data satisfy   
Einstein's constraint equations. 
Then the corresponding initial value problem for the Einstein equations admits a globally hyperbolic Cauchy development $(M,g)$, which is also endowed with a global foliation by asymptotically flat hypersurfaces, 
is future causally geodesically complete, and is asymptotic to Minkowski spacetime in future causal directions, as well as 
in spacelike directions.

\vskip-.cm
\begin{figure}
\hskip2.6cm 
  \begin{picture}(0,0)
  \put(18,-108){\includegraphics[width=6cm,height=3.5cm]{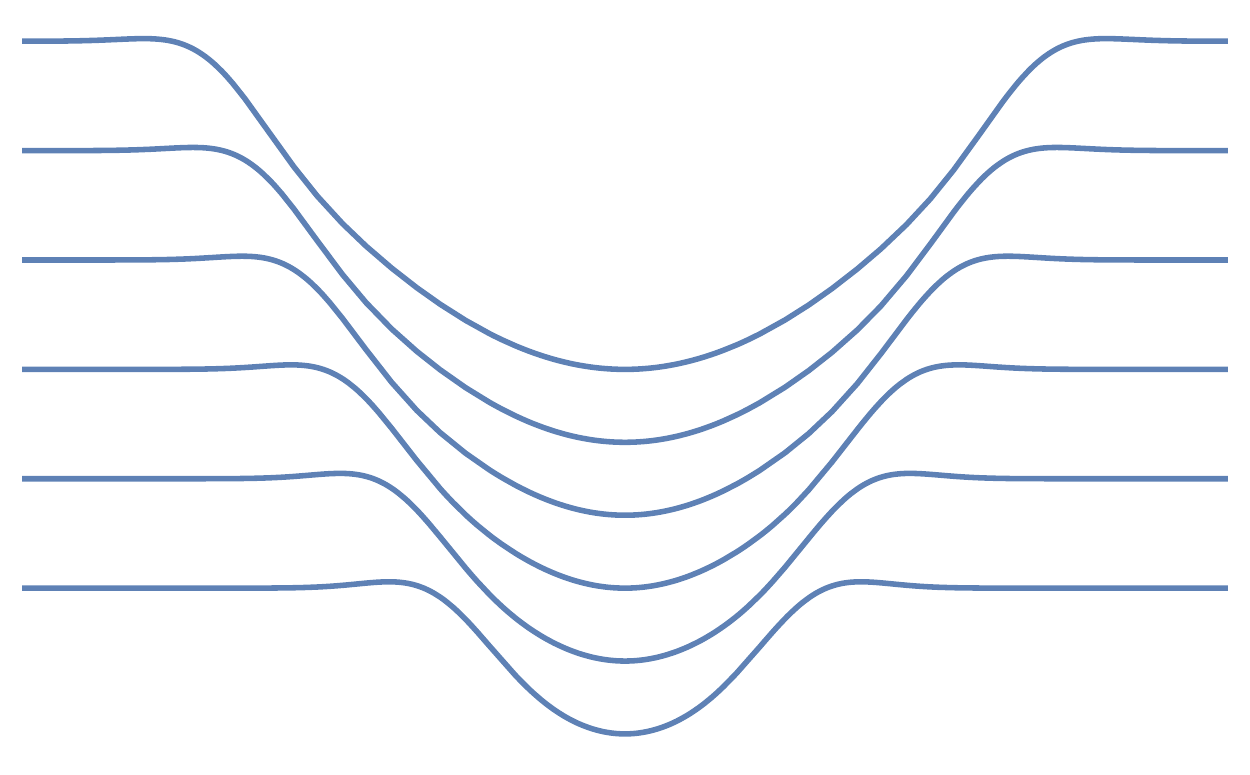}}
  \put(92,-35){asympt.}  
  \put(92,-50){hyperb.}
    \put(130,-93.9){asympt. Euclidian}  
\end{picture}
\vskip4.cm
\caption{\label{figure1}The Euclidian-hyperboloidal foliation} 
\end{figure}
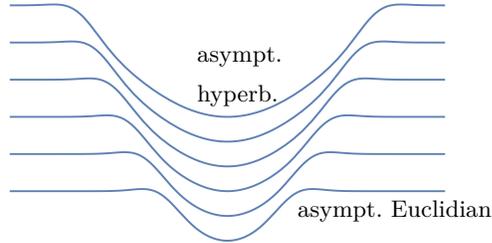

\paragraph{The Euclidian-hyperboloidal foliation method.}

Following the pioneering work by Lindblad and Rodnianski~\cite{LR-2}, 
we use the so-called wave gauge, and we thus introduce global coordinate functions satisfying the wave equation in the unknown metric. 
Our proof is based on a new methodology of proof which combines several novel ideas. We construct a  
foliation consisting 
of 
\bei

\item (1) asymptotically hyperboloidal slices in the interior of a light cone (which plays an important role in deriving decay in time), 

\item 
(2) asymptotically Euclidian slices in the exterior of this light cone (which plays an important role in deriving decay in space), 
and 

\item 

(3) we merge these two foliations in the vicinity of the light cone.  

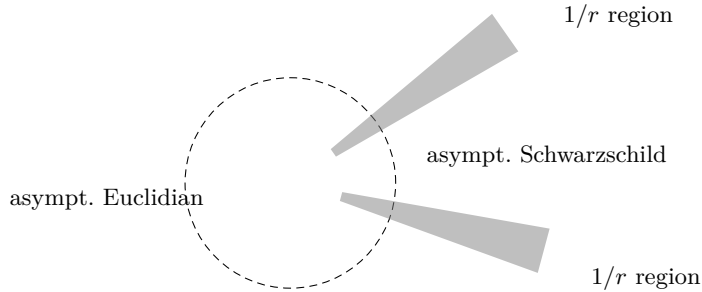
\begin{figure}
\hskip1.5cm 
\begin{picture}(0,0)
\put(0,0)
{
\begin{tikzpicture}[scale=.7]
  \draw[densely dashed] (0,0) circle (2);
    \fill[fill=gray, opacity=.5](30:1) -- (30:5) -- (40:5) -- (40:1) -- cycle;
  \node at (27:6.99) { $1/r$ region};
      \fill[fill=gray, opacity=.5](-10:1) -- (-10:5) -- (-20:5) -- (-20:1) -- cycle;
  \node at (-15:6.99) { $1/r$ region};
  \node at (185:3.5) { asympt.\ Euclidian};
  \node at (6:4.9) { asympt.\ Schwarzschild};
\end{tikzpicture}
}
\end{picture}
\vskip-.cm
\caption{\label{figure2}Asymptotic behavior at infinity.} 
\end{figure}

\eei
\noindent 
{  See the illustration in  in Fig.~\ref{figure1}.} 
Moreover, the behavior in spacelike directions can be rather general; for instance the metric may be asymptotic 
 with the Minkowski metric and the Schwarzschild metric in certain angular directions (with the exception of a cone with arbitrarily small angle where it may still enjoy $1/r$ decay \cite{LeFlochNguyen}. 
{ 
See the illustration in Fig.~\ref{figure2}. 
}
We also rely on (approximate) symmetries associated with the geometry of Minkowski spacetime but, importantly, we avoid the use of   the scaling field, since it does not commute with the Klein-Gordon operator. We derive  
sharp energy estimates as well as sharp pointwise estimates for both the geometric and the matter variables. This requires us to establish new (weighted) Sobolev, Hardy, and Poincar\'e inequalities.  Furthermore, we carefully analyze the nonlinear coupling taking place between the geometry and matter. More generally, our method applies to the global existence problem for a broad class of 
coupled nonlinear wave-Klein-Gordon equations.


\vskip4.cm

\section{Spacetimes with singularity hypersurfaces} 
\label{section--3} 

\subsection{Our standpoint}
\label{section--31} 

\paragraph{Asymptotics at singularities.}

We now present our framework to analyze spacetimes with singularities, 
and we seek a flexible framework for bouncing cosmologies involving contracting/expan\-ding evolution phases. We are interested in covering physically meaningful junction conditions and, as mentioned earlier,
 we must go beyond Israel's standard junction conditions since they 
only apply to regularity hypersurfaces. Here, we outline our new methodology and main results, 
while referring to the main papers for full statements and proofs~\cite{LLV-1a,LLV-2,LLV-1b}. 

In the regime of interest, we may encounter a rich and complex dynamics near singularities 
and we consider spacetimes in the quiescent regime which, by definition, enjoy certain monotone behavior 
and the absence of BKL oscillations (after Belinsky, Khalatnikov, and Lifshitz). 
Such spacetimes in the quiescent regime admit asymptotic expansions near a singularity hypersurface, and 
are found in ta large variety of setups, including in the description of self-gravitating scalar field, stiff fluid, 
or compressible fluid, as well as all (matter or vacuum) spacetimes with symmetries (for instance, spatial $T^2$ symmetry). 

\paragraph{Methodology.}

We work with suitably notions of rescaled metric, intrinsic curvature tensor, and matter fields.
For the description of the geometry near a singularity hypersurface we require 

\bei  

\item a set of past and future {singularity data} denoted by $(g^\pm, K^\pm, \phi_0^\pm, \phi_1^\pm)$, and    

\item a {singularity scattering map} denoted by 
$$
\Sbf:(\gmoi, \Kmoi, \phimoi_0, \phimoi_1) \mapsto (g^+, K^+, \phi_0^+, \phi_1^+).
$$
\eei
\noindent 
In turn, based on these notions and in a sense we introduce, we end up constructing spacetimes ``beyond'' singularities and, in the case of plane-symmetry, globally-defined $S$-cyclic spacetimes. 

A first issue is to parametrize the degrees of freedom associated with the constraints at the singularity, 
while the second main issue is classifying the set of all possible bouncing laws.
Interestingly, thanks to a systematic study of the set of singularity scattering maps we arrive at a general classification,
 by working with general spacetimes and relying on Einstein's constraint equations. 
 Observe that all earlier works in this question
considered symmetric spacetimes or special junction conditions and, therefore, provided only a partial view on the problem.  
{ 
On the other hand, our work leads to a complete characterization of all physically-relevant maps, which clearly 
separates between universal and model-dependent features of the scattering on gravitational singularities.
}


\subsection{Formulation of the problem} 
\label{section--32} 

\paragraph{The local ADM formulation.}

For simplicity in the present review we restrict attention to spacelike hypersurfaces. 
We introduce a local ADM formulation near the singularity hypersurface under consideration, which consists of 
a Gaussian foliation covering a (small neighborhood in a) spacetime 
 by spacelike hypersurfaces diffeomorphic to a given slice $\Hcal_0$, say 
$$
\Mcal^{(4)} = \bigcup_{\tau \in [\tau_{-1}, \tau_1]} 
 \Hcal_\tau,
$$
together with a spacetime metric 
$$
g^{(4)} = \big(g^{(4)}_{\alpha\beta}\big)
= - d\tau^2 + g(\tau), 
\qquad  
 g(\tau) = g_{ij}(\tau) dx^i dx^j. 
$$
in which $\tau$ remains in a neighborhood of the origin, that is, $0 \in [\tau_{-1}, \tau_1]$. 
 
We require that Einstein's evolution equations for the  induced metric $g$ and the intrinsic curvature $K$ hold, that is,  
\be
\del_\tau g_{ij}  = - 2 \, K_{ij}, 
\qquad  
\del_\tau K^i_j = \Tr(K)  K^i_j + R^i_j - 8 \pi \, M^i_j. 
\ee
Here $M^i_j = {1 \over 2} \rho \, g^i_j + T^i_j - {1 \over 2} \Tr(T)  g^i_j$ denotes the matter contribution
corresponding to a matter field $\phi$, typically satisfying the wave equation 
\be
\Box_{g^{(4)}} \phi = 0.  
\ee 
The formulation is supplemented with Einstein's constraint equations (Hamiltonian, momentum) 
\be
R + |K|^2 - \Tr (K^2) = 16 \pi \rho, 
\qquad 
\nabla_i K^i_j  - \nabla_j (\Tr K) = 8 \pi J_j. 
\ee  
 
When the time function is chosen to be such that the slices have constant mean curvature, we obtain the so-called CMC-Einstein flow which was studied extensively away from singularities by Andersson and Moncrief (local existence theory) and 
Anderson, Lott, Moncrief, Reiris, etc. (global dynamics theory). 
In the present discussion, we are interested in the {\sl local behavior near a singularity hypersurface.}

 
\paragraph{Example of asymptotic behavior:  Kasner profiles.} 

As a first illustration of the asymptotic behavior that one should expect, let us consider the metric, extrinsic curvature, and matter field given by { (with $\tau \in (-1,0)$)} 
\be
\aligned
g^*_\Kasner(\tau,x) 
& = (- \tau)^{2p_1(x)} (dx^1)^2 + (- \tau)^{2p_2(x)} (dx^2)^2 +  (- \tau)^{2p_3(x)} (dx^3)^2, 
\\
K^*_\Kasner(\tau,x) & = {-1 \over \tau} \diag(p_1,p_2,p_3)(x)
\\
\phi^*_\Kasner(\tau,x) & = \phimoi_0(x) \log|\tau|  + \phimoi_1(x). 
\endaligned
\ee
Here, the Euclidean metric $\gmoi$ on $\Hcal \simeq \RR^3$ is chosen, while $\Kmoi$ has constant eigenvectors
 and $\Kmoi \equiv \diag(p_1,p_2,p_3)$ in suitable coordinates. The
  functions $p_1, p_2, p_3$ are prescribed and defined on~$\RR^3$. In addition, we 
  choose  matter data $(\phimoi_0,\phimoi_1)$  which are also $x$-dependent. 

Under suitable conditions on the data  $p_1, p_2, p_3$ and $(\phimoi_0,\phimoi_1)$, this is an ``asymptotic profile'' 
{ in the limit $\tau \to 0$}, 
in the sense we define next. We can also introduce the generalized Kasner spacetime metric 
\be
g_\Kasner^{*\,(4)} = - d\tau^2 + g^*_\Kasner(\tau).  
\ee
We distinguish between the particular cases. 
 
\bei  

\item { Case $\phimoi_1$ constant:} we then have 
$
p_j(x) = 1/3 + f_{j+1}(x^2) - f_{j+2}(x^3)  
$, 
which is 
parametrized by three functions on~$\RR$, subject only to the inequality
\be
\sum_j p_j(x)^2 \leq 1,
\ee
 easily satisfied for example by functions with all $\abs{f_j(x^j)}$ sufficiently small. { This is only an asymptotic solution (and generally not an exact solution) to the 
 Einstein-scalar field system.} 

\item { Case $p_j$ constant:} this is Kasner spacetime, not just an asymptotic profile, but the well-known solution to the (matter) Einstein equations. It is a vacuum solution only if, moreover, $\phi_0^-$ vanishes.
 
\eei

 
\section{Fundamental notions and local existence theory} 
\label{section--4} 

\subsection{A construction scheme} 

Motivated by the work by Rendall \cite{Rendall:2008} and followers, we adopt the following strategy 
{ in order to parametrize a class of bouncing spacetimes. } 
\bei

\item[--] We make a gauge choice ensuring that the singularity hypersurface is located at $\tau=0$.  

\item[--] We solve from $\tau=0$ toward the past ($\tau<0$) and toward the future ($\tau>0$). 

\item[--] We derive an asymptotic ODE system referred to as the  ``velocity dominated'' Ansatz'' which consists of 
(essentially) removing all spatial dependency in a choice of local coordinates. 

\item[--] We apply argument from the theory of Fuchsian equations (Baouendi, Goul\-aouic, Rendall, Kichenassamy,\ldots).  

\item[--] We check that the asymptotic data and the asymptotic constraints hold on the singularity hypersurface.  

\eei 

\noindent In addition, we glue together past and future solutions, using a junction condition. 


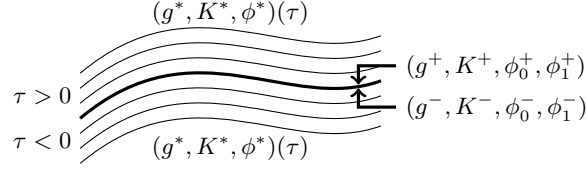
\begin{figure}\centering
  \begin{tikzpicture}
    \draw[very thick] (-2,0) .. controls +(40:2) and +(-160:1.3) .. +(4,.5);
    \foreach \y in {-.6,-.4,-.2,.2,.4,.6} {
      \draw (-2,\y) .. controls +(40:2) and +(-160:1.3) .. +(4,.5); }
    \node at (-2,.3) [left] {$\tau>0$};
    \node at (-2,-.3) [left] {$\tau<0$};
    \node at (0,1.4) {$(g^*,K^*,\phi^*)(\tau)$};
    \draw [->, very thick] (2.2,.7) node [right] {$(\gpoi,\Kpoi,\phipoi_0,\phipoi_1)$} -- (1.7,.7) -- (1.7,.45);
    \draw [->, very thick] (2.2,.15) node [right] {$(\gmoi,\Kmoi,\phimoi_0,\phimoi_1)$} -- (1.7,.15) -- (1.7,.4);
    \node at (0,-.4) {$(g^*,K^*,\phi^*)(\tau)$};
  \end{tikzpicture}
\caption{\label{figure3}Spacetime foliation by spacelike hypersurfaces~$\Hcal_\tau$.} 
\end{figure}

\subsection{Singularity data and asymptotic profiles}
\label{section--36} 

We propose the following notions. See the illustration in Fig.~\ref{figure3}. 

\begin{definition}
 A {\bf (past) singularity initial data set} $(\gmoi, \Kmoi, \phimoi_0, \phimoi_1)$ defined on a $3$-manifold $\Hcal$ consists of  
 two tensor fields $(\gmoi, \Kmoi)$ and two scalar fields $(\phimoi_0, \phimoi_1)$ such that:  
 \bei 
\item[(i)] $\gmoi  = (g^-_{ij})$ is a Riemannian metric  on $\Hcal$. 

\item[(ii)] $ \Kmoi = (K_i^{-j})$ is a CMC symmetric $(1,1)$-tensor, namely satisfying 
\be
\gmoi_{ik} \Kmoi_j^k = \gmoi_{jk} \Kmoi_i^k.
\ee

\item[(iii)] The constant mean-curvature condition $\textbf{Tr}(\Kmoi) = 1$ holds on $\Hcal$. 

\item[(iv)] The Hamiltonian constraint holds 
\be
 1 - | K^-|^2
= 8\pi \, (\phimoi_0)^2. 
\ee

\item[(v)] The momentum constraints hold 
\be
\textbf{div}_{g^-} (K^-)
= 8 \pi \, \phimoi_0 d\phimoi_1. 
\ee

\eei 
Furthermore, the collection of all singularity data sets is denoted by  $\Ibf(\Hcal)$. 
\end{definition}

\begin{definition}
A {\bf (past) asymptotic profile}  associated with a singularity initial data set 
$(\gmoi, \Kmoi, \phimoi_0, \phimoi_1) \in \Ibf(\Hcal)$ is the 
following ancient geometric flow defined on $\Hcal$ 
\be
\tau \in (-\infty,0) \mapsto \big(g^*, K^*, \phi^*\big) (\tau) 
\ee
{
as follows (with the exponential notation $|\tau|^{2 \Kmoi} = e^{2 \log(|\tau|) \, \Kmoi}$):
}
\be
\aligned  
g^*(\tau) & = |\tau|^{2 \Kmoi} \gmoi,  
\\ 
K^*(\tau) &
 = {-1 \over \tau} \Kmoi, 
\\ 
\phi^*(\tau) & = \phimoi_0 \log|\tau|  + \phimoi_1.
\endaligned
\ee
\end{definition}

The regime of interest in the present work corresponds to the so-called 
quiescent singularities having
\be
 K^->0. 
 \ee
Below, we will also require the same sign condition after the bounce (``tame preserving''). In particular, this condition easily implies that 
the volume element decreases to zero as $\tau\to 0^-$,  and 
  then increases back to finite values for $\tau>0$, as should be expected for a bounce.


\subsection{Cyclic spacetimes}  
\label{section--33} 

Our novel concepts are as follows.

\begin{definition}
A {\bf past-to-future singularity scattering map} on a manifold $\Hcal$ by definition is a map
\be
\Sbf: \Ibf(\Hcal) \to \Ibf(\Hcal), 
\qquad 
(\gmoi, \Kmoi, \phimoi_0, \phimoi_1) \mapsto (\gpoi, \Kpoi, \phipoi_0, \phipoi_1)
\ee
satisfying the two conditions: 
\bei 
 
\item {\bf Diffeomorphism-covariance}, that is, coordinate invariance. 
 
\item {\bf Locality property}, that is,  the restriction of $\Sbf(\gmoi, \Kmoi, \phimoi_0, \phimoi_1)\big|_U$
 depends only on the restriction of the data,  for any open set $U \subset \Hcal$.
 
\eei 

\end{definition}

\begin{definition}
Fix a singularity scattering map $\Sbf$. A {\bf $\Sbf$-cyclic spacetime} $(\Mcal^4, g)$ by definition satisfies the following conditions: 

\bei 

\item $\Mcal^4$ is a manifold endowed with a Lorentzian metric~$g^{(4)}$ and a scalar field~$\phi$.

\item {\sl Regularity domain:} $g^{(4)}$ and $\phi$ are defined outside a {\bf singularity locus} 
 $\Lcal \subset \Mcal^4$, 
and
{ 
 the Einstein equations hold  in $\Mcal^4 \setminus \Lcal$, 
 }
 that is, the 
(evolution and constraint) Einstein equations $G^{(4)}_{\alpha\beta}= 8 \pi T^{(4)}_{\alpha\beta}$ 
together with the matter evolution equation $\Box_{g^{(4)}} \phi = 0$ (which actually follows from the former).

\item {\sl Local Gaussian foliations:} 
every point $p \in \Lcal$ admits a neighborhood~$\Ucal$  endowed with 
 a foliation  
  $\bigcup_{\tau} \Hcal_\tau$ containing $\tau=0$ and such that $\Hcal_0= \Lcal \cap \Ucal$.
  For $\tau\neq 0$, $\Hcal_\tau$ are spacelike and diffeomorphic to~$\Hcal_0$
and the $4$-metric reads $g^{(4)}=-d\tau^2+g^{(3)}(\tau)$ for some $3$-metrics $g(\tau)$
  defined on~$\Hcal_\tau \simeq \Hcal_0$.

\item {\sl Junction conditions on $\Lcal$}: the future and past singularity data 
\be
(g^\pm, K^\pm, \phi^\pm_0, \phi^\pm_1)
= 
\lim_{\tau \to 0 \atop \tau \gtrless 0} 
\big(|\tau|^{2 \tau K} g,\; -\tau K,\; \tau \del_\tau \phi,\; \phi - \tau \log |\tau|  \del_\tau \phi \big)(\tau)
\ee
are related by the relations 
\be
(\gpoi, \Kpoi, \phipoi_0, \phipoi_1) = \Sbf(\gmoi, \Kmoi, \phimoi_0, \phimoi_1).
\ee

\eei 
\end{definition} 

   
 \vskip3.5cm
 
 \begin{figure}
 \hskip3.cm 
 \begin{picture}(0,0)
\put(0,0)
{ 
  \begin{tikzpicture}  [scale=.99]  
    \draw[very thick] (-2,-.2)
    .. controls +(-40:.4) and +(-160:.6) .. (-.7,-.3)
    .. controls +(20:.6) and +(170:.5) .. (.7,0)
    .. controls +(-10:.6) and +(180:.5) .. (2,0);
    \foreach \y in {-.7,-.6,-.5,-.4,-.2,-.1,0,.1} {
      \draw (-.7,\y) .. controls +(20:.6) and +(170:.5) .. +(1.4,.3); }
    \draw (-.7,-.7) -- (-.7,.1);
    \draw (.7,-.4) -- (.7,.4);
    \node at (0,.7) {};
    \node at (0,-.065) {$\bullet$};
    \draw[-{stealth}, thick] (.5,-.7) node [right] {$p$} -- (0,-.7) -- (0,-.15);
    \node at (1.5,.3) {$\tau>0$};
    \node at (1.5,-.3) {$\tau<0$};
    \node at (-2.2,0) { $\Lcal$};
    \draw[ dashed] (-2,1) .. controls +(40:2) and +(-160:1.3) .. (2,1.5);
    \draw[ dashed] (-2,-1) .. controls +(-40:1.3) and +(145:1.4) .. (2,-2);
  \end{tikzpicture}
}
\end{picture}
 \caption{\label{fig:singu2}Cyclic universe with singular locus denoted by~$\Lcal$,
  consisting of three singularity hypersurfaces.} 
\end{figure}
 

\subsection{Existence and asymptotic properties of cyclic spacetimes}

We also say that a scattering map is  {\bf quiescence-preserving} provided $K^+ >0$ whenever $K^- >0$, 
that is, the map preserves the positivity of the intrinsic curvature. 
We then arrive at our existence result which we state somewhat informally. For more precise statements we refer the reader to \cite{LLV-1a}.

\begin{theorem}[Existence of a class of cyclic spacetimes. The glueing technique]
Consider a three-manifold $\Hcal_0$ and quiescence-preserving scattering map $\Sbf: \Ibf(\Hcal_0) \mapsto \Ibf(\Hcal_0)$ 
 given over the space of singularity data. 
Consider a quiescent singularity data $(\gmoi, \Kmoi, \phimoi_0, \phimoi_1)$ defined on $\Hcal_0$, that is, 
satisfying the positivity condition $\Kmoi > 0$. 
Then  there exists a $\Sbf$-cyclic spacetime $(\Mcal^{(4)},g^{(4)})$ with singularity locus $\Hcal_0$, 
together with a locally Gaussian foliation $\Mcal^{(4)}=\bigcup_{\tau \in [\tau_{-1}, \tau_1]} \Hcal_\tau$ with time coordinate~$\tau$,
such that the flow $\tau \mapsto (g(\tau), K(\tau), \phi(\tau))$ satisfies the Einstein equations coupled to a scalar field $\phi$ away from $\tau=0$,
and $(\gpoi, \Kpoi, \phipoi_0, \phipoi_1) = \Sbf(\gmoi, \Kmoi, \phimoi_0, \phimoi_1)$ holds on $\Hcal_0$.

If $\Hcal_0$ is compact, then the volume $V(\tau) = \Vol_{g(\tau)}(\Hcal_\tau)$ of the slices  
is shrinking toward the singularity
\be
\lim_{\tau \to 0}  V(\tau) = 0.
\ee 
The solution exhibits a {\rm crushing singularity} in the sense that the  mean curvature of the slices
blowup 
\be
\lim_{\tau \to 0} \tau H(\tau) = -1 \quad \text{ on } \Hcal_\tau
\ee
Moreover, the solution exhibits a {\rm curvature singularity} at which the spacetime scalar (and Weyl) curvature 
 $R^{(4)}$ blows up  in a uniform way on the singularity hypersurface:  
\be
\lim_{\tau \to 0^{\pm}} \tau^2 R^{(4)}(\tau)  =  - 8 \pi (\phi_0^{\pm})^2 
\, \text{ on } \Hcal_\tau
\ee
as well as the spacetime Weyl curvature except in degenerate cases 
\end{theorem} 
 
As mentioned earlier, our technique of proof is of Fuchsian type and relies on a glueing argument
For additional developments on Fuchsian techniques and related issues, we refer to standard papers by 
Rendall and co-authors \cite{Rendall:2008,RendallWeaver} and more recent contributions by Alexakis, 
Fournodavlos, Luk, Speck, and Rodnianski, referred to in \cite{LLV-1a}.


\section{Classification of scattering maps}
\label{section--5} 

\subsection{Terminology}
\label{section--41}

We continue with some further definitions.  

\begin{definition} A singularity scattering map $\Sbf$ is said   to enjoy the {\bf locality property} if,  
 for all point $x \in \Hcal$,  
$\Sbf\big(\gmoi, \Kmoi, \phimoi_0, \phimoi_1\big)(x)$ depends upon  
$\big(\gmoi, \Kmoi, \phimoi_0, \phimoi_1\big)(x)$ 
and possibly derivatives at $x$, only. 

It is called  a {\bf ultra-local map} if   it involves pointwise values only, that is, 
$\Sbf(\gmoi, \Kmoi, \phimoi_0, \phimoi_1)(x)$
depends only on $(\gmoi, \Kmoi, \phimoi_0, \phimoi_1)(x)$. By diffeomorphism invariance, the restrictions $\Sbf_x$ to every point~$x$ then are the same.

A {\bf conformal map} by definition is such that $g^*(\tau_-)$ and $g^*(\tau_+)$ differ by a conformal factor. A map is said to be 
 {\bf rigidly conformal} if  
$\gpoi$ and $\gmoi$ differ by a conformal factor
\end{definition}

Due to the ultralocality property, specifying a singularity scattering map~$\Sbf$ on~$\Hcal$ is equivalent to specifying one on a unit ball of $\Hcal$. 

\begin{definition}  A singularity scattering map $\Sbf$ 
is said to be
\bei 

\item {\bf momentum-preserving} if   $K^+ = K^-$ and  $\phipoi_0=\pm\phimoi_0$; 

\item {\bf momentum-reversing} if  $\Kpoi=\frac{2}{3}\delta-\Kmoi$ and $\phipoi_0=\pm\phimoi_0$; 

\item  {\bf idempotent} if  
$\Sbf \circ \Sbf$ is the identity map on $\Ibf(\Hcal)$; 

\item {\bf invertible} if   $\Sbf^{-1}$ is well-defined as a scattering map. 

\eei 
{ 
(Here, we denote by $\delta$ the Kronecker symbol $\delta_i^j$.)
}
\end{definition}

  
\subsection{Main classification results}
\label{section--42} 

Finally, we are in a position to state our main results. 
 
\begin{theorem}[Rigidly conformal maps] 
Only two classes of ultra-local spacelike rigidly conformal singularity scattering maps are available 
for bouncing of self-gravitating scalar fields. They are described as follows.  
  
\bei 

\item {\bf Isotropic rigidly conformal bounce} $\Sbf^\text{iso}_{\lambda,\varphi}$: 
\be
g^+ = \lambda^2 \gmoi,
\qquad
K^+ = \delta/3,
\qquad
\phi_0^+ =  1/\sqrt{12\pi},
\qquad
\phi_1^+ =  \varphi,
\ee
parametrized by a conformal factor $\lambda=\lambda(\phimoi_0,\phimoi_1,\det\Kmoi)>0$  and a constant~$\varphi$.


\item  {\bf Non-isotropic rigidly  
conformal bounce}  $\Sbf^\text{ani}_{f,c}$: 
\be
\aligned
&  g^+ = c^2 \mu^2 \gmoi,
\qquad
&&
K^+ = \mu^{-3}(\Kmoi - \delta/3) + \delta/3,
\\
& 
\phi_0^+ =  \mu^{-3} \phimoi_0/F_\phi(\phimoi_1),
\qquad
&&
\phi_1^+ = F(\phimoi_1),
\endaligned
\ee
parametrized by a constant ${c>0}$ and a function $f\colon\RR\to [0, +\infty)$
\be
\mu(\phi_0, \phi_1)  
=  \big(1+12\pi (\phi_0)^2 f(\phi_1) \big)^{1/6}, 
\qquad 
F(\phi_1)= \int_0^{\phi_1} (1+f(\varphi))^{-1/2} d\varphi. 
\ee

\eei 
\end{theorem} 

An analogous result holds for general maps; see~\cite{LLV-1a} for the full statement.  
 
\begin{theorem}[General classification] 
Only two classes of ultra-local spacelike singularity scattering maps for self-gravitating scalar fields, 
{ which 
represent either 
 an isotropic bounce denoted by $\Sbf^\text{iso}_{\lambda,\varphi}$  or
 a non-isotropic bounce  denoted by $\Sbf^\text{ani}_{\Phi,c}$.
 Now, 
 }
 $\lambda$ is a two-tensor, $\Phi$ is a canonical transformation, and $c$ is a constant.  
\end{theorem} 
 

\subsection{The three universal laws of quiescent bouncing cosmology} 

We complete our presentation by stating three universal laws obeyed by {\sl any} ultra-local bounce.

\bei 

\item {\bf First law: scaling of Kasner exponents}.  There exists a (dissipation) constant $\gamma\in\RR$ such that
\be
 |g^+ |^{1/2} \Kcirc^+ = - \gamma \, |g^- |^{1/2} \Kcirc^-
\ee
in which we recall that $g$ denotes the (rescaled) spatial metric in synchronous gauge,
$|g|^{1/2}$ is the corresponding  volume factor, and $\Kcirc$ denotes the traceless part 
of the extrinsic curvature (as a $(1,1)$ tensor).

\item {\bf Second law: canonical transformation of matter}. There exists a nonlinear map 
$\Phi\colon(\pi_\phi,\phi)^- \mapsto (\pi_\phi,\phi)^+$, 
preserving the volume form in the phase space $d\pi_\phi\wedge d\phi$
and depending solely on the scalar invariant $\textbf{det}(\Kcirc_-)$. 
Here,   $\pi_\phi \sim \phi_0$ denotes the conjugate momentum. 
 
\item {\bf Third law: directional metric scaling}. One has 
$$
\aligned
& g^+= e^{(\sigma_0 + \sigma_1 K + \sigma_2 K^2)} \, g^-, 
\endaligned
$$ 
which involves a nonlinear scaling in each proper direction of~$K$. 
When $\gamma=0$, we have an isotropic scattering and no restriction on $\sigma_0,\sigma_1,\sigma_2$. 
When $\gamma\neq 0$, one has a non-isotropic scattering, explicit formulas for 
$\sigma_0,\sigma_1,\sigma_2$ are available in terms of $\Phi,\gamma$.

\eei


\subsection{Role of the small-scale physics} 
\label{section--43} 
 
 Let us conclude this overview with some further references. 
 In ongoing work, we are interested in deriving scattering maps associated with specific theories. 
Our methodology encompasses junction conditions that were proposed in a variety of contexts: 
pre-Big Bang scenario (Gasperini, Veneziano, etc.),
 modified gravity-matter models (Brandenberger, Peter, Steinhardt, Turok, etc.), 
 and  loop quantum cosmology (Asthekar, Pawlowski, Wilson-Ewing, etc.) among other theories.  
 A different proposal was also made by Penrose \cite{PenroseCCC1} and studied by Tod \cite{Tod} and followers. 

Interestingly, as pointed out in \cite{LeFlochLeFloch-3} the notion of singularity scattering map 
naturally connects with the notion of ``kinetic relation'' that was proposed 
for sharp interface models of phase transition dynamics, for instance 
for two-phase flows of fluids or elastic materials. In this context \cite{LeFloch-Oslo}, the small-scale parameters of interest,
 which can be neglected at the macro-scale level of description,
 account for the viscosity, surface tension, heat conduction, etc.  
The essential macro-scale features  are captured by junction conditions of 
Rankine-Hugoniot type, as well as
kinetic relations or DLM families of paths (after Dal~Maso, LeFloch, and Murat); see \cite{LeFloch-IMA}. 

 In turn, we have proposed a flexible framework for dealing with junction conditions in general relativity. 
 We have uncovered all possible classes of junctions that are both geometrically and physically meaningful, and 
 our classification distinguish between 
junctions of conformal or non-conformal type, spacelike or null or timelike type, etc. 
 The methodology applies to scalar fields as well as stiff fluids or compressible fluids, and provides us
 with a guide in order to uncover relevant structures for each model of interest. In particular, our three universal laws
   constrain the  macroscopic aspects of spacetime bounces, regardless of their origin from different microscopic corrections 



\begin{thebibliography}{6}


\bibitem{AnderssonRendall}
\auth{L.~Andersson and A.D.~Rendall,}
Quiescent cosmological singularities,
Commun.\ Math.\ Phys.\ 218 (2001), 479--511.

\bibitem{Ashtekar}
\auth{A.~Ashtekar,}
Singularity resolution in loop quantum cosmology: a brief overview,
J.~Phys.\ Conf.\ Ser.\ 189 (2009), 012003.

\bibitem{AshtekarWilsonEwing}
\auth{A.~Ashtekar and E.~Wilson-Ewing,}
Loop quantum cosmology of Bianchi I models,
Phys.\ Rev.\ D 79 (2009), 083535.

\bibitem{BarrabesIsrael}
\auth{C.  Barrab\`es and W. Israel,}
Thin shells in general relativity and cosmology: the lightlike limit, 
Phys. Rev. D 43 (1991), 1129--1142.

\bibitem{Barrow}
\auth{J.D.~Barrow,}
Quiescent cosmology,
Nature 272 (1978), 211--215.

\bibitem{BV} 
\auth{V.~Bozza  and  G.~Veneziano,}
Regular two-component bouncing cosmologies and perturbations therein,
J.\ Cosmol.\ Astro.\ Phys.\  9 (2005), 007.

\bibitem{Brizuela:2009nk}
\auth{D.~Brizuela, G.A.D.~Mena Marugan, and T.~Pawlowski,}
Big Bounce and inhomogeneities,
Class.\ Quant.\ Grav.\ 27 (2010), 052001.

\bibitem{FloresSanchez:2003}
\auth{J.L. Flores and M. S\'anchez,}
Causality and conjugate points in general plane waves,
Class. Quantum Grav. 20 (2003) 2275--2291.

\bibitem{FloresSanchez:2008}
\auth{J.L. Flores and M. S\'anchez,}
The causal boundary of wave-type spacetimes, 
J. High Energy Phys. (2008), 036.

\bibitem{IonescuPausader0} 
{\sc A.D. Ionescu and B. Pausader,} 
On the global regularity for a wave-Klein-Gordon coupled system, 
Acta Math. Sin. 35 (2019), 933--986. 

\bibitem{IonescuPausader} 
{\sc A.D. Ionescu and B. Pausader,} 
The Einstein-Klein-Gordon coupled system: global stability of the Minkowski solution, 
Preprint arXiv:1911.10652.  

\bibitem{LeFlochLeFloch-1}
\auth{B.~Le Floch and P.G.~LeFloch,}
On the global evolution of self-gravitating matter. Nonlinear interactions in Gowdy symmetry,
Arch.\ Rational Mech.\ Anal.\  233 (2019), 45--86.

\bibitem{LeFlochLeFloch-2}
\auth{B.~Le Floch and P.G.~LeFloch,} 
Compensated compactness and corrector stress tensor for the Einstein equations in $\Tbb^2$ symmetry, 
{Portugaliae Math.} 77 (2020), 409--421. See also ArXiv:1912.12981. 

\bibitem{LeFlochLeFloch-3}
\auth{B.~Le Floch and P.G.~LeFloch,}
On the global evolution of self-gravitating matter. Scattering maps for interfaces, 
 in preparation. 

\bibitem{LeFlochLeFloch-4}
\auth{B.~Le Floch and P.G.~LeFloch,}
On the global evolution of self-gravitating matter. $\Tbb^2$ areal flows and compensated compactness, 
 in preparation.

\bibitem{LLV-1a}
\auth{B.~Le Floch, P.G.~LeFloch, and G.~Veneziano,}
Cyclic spacetimes through singularity scattering maps. 
The laws of bouncing cosmology, 
Preprint ArXiv:2005.11324.

\bibitem{LLV-2}
\auth{B.~Le Floch, P.G.~LeFloch, and G.~Veneziano,}
Universal scattering laws for quiescent bouncing cosmology, 
Physical Rev. D 8 (2021), 083531. See also ArXiv:2006.08620. 

\bibitem{LLV-1b}
\auth{B.~Le Floch, P.G.~LeFloch, and G.~Veneziano,}
Cyclic spacetimes through singularity scattering maps. 
Plane-symmetric collisions, Preprint ArXiv:2106.09666. 

\bibitem{LeFloch-IMA}
\auth{P.G.~LeFloch,}
Shock waves for nonlinear hyperbolic systems in nonconservative form, 
Institute for Math. and its Appl., Minneapolis, IMA, Preprint \# 593, 1989. 
Available at: 
\\
https://conservancy.umn.edu/bitstream/handle/11299/5107/593.pdf 

\bibitem{LeFloch-Oslo}
\auth{P.G.~LeFloch,}
Kinetic relations for undercompressive shock waves. Physical, mathematical, and numerical issues, 
Contemp.\ Math.\ 526 (2010), 237--272.
 
\bibitem{LeFlochMa1} {\sc P.G. LeFloch and Y. Ma,}
{\sl The hyperboloidal foliation method,}  World Scientific Press, 2014.
 
\bibitem{LeFlochMa2}{\sc P.G. LeFloch and Y. Ma},
The global nonlinear stability of Minkowski space for self-gravitating massive fields. The wave-Klein-Gordon model,
Comm. Math. Phys. 346 (2016), 603--665.
 
\bibitem{LeFlochMa3}{\sc P.G. LeFloch and Y. Ma},
{\sl The global nonlinear stability of Minkowski space for self-gravitating massive fields,} 
World Scientific Press, 2018. 
 
\bibitem{LeFlochMa4}{\sc P.G. LeFloch and Y. Ma},
Nonlinear stability of self-gravitating massive fields, Preprint ArXiv:171210045, submitted for publication.  


\bibitem{PLFMardare} 
\auth{P.G.~LeFloch and C.~Mardare,}
Definition and weak stability of spacetimes with distributional curvature, 
Portugal Math.\ 64 (2007), 535--573.

\bibitem{LeFlochNguyen} 
\auth{P.G.~LeFloch and T.-C. Nguyen,} 
The seed-to-solution method for the Einstein equations and the asymptotic localization problem, 
Preprint ArXiv:1903.00243.

\bibitem{PLFRendall}
\auth{P.G.~LeFloch and A.D.~Rendall,}
A global foliation of Einstein-Euler spacetimes with Gowdy-symmetry on $T^3$,
Arch.\ Rational Mech.\ Anal.\ 201 (2011), 841--870.

\bibitem{PLFSormani}
\auth{P.G.~LeFloch and C.~Sormani,}
The nonlinear stability of rotationally symmetric spaces with low regularity,
J.\ Funct.\ Anal.\ 268 (2015), 2005--2065.

\bibitem{PLFStewart}
\auth{P.G.~LeFloch and J.M.~Stewart,}
The characteristic initial value problem for plane--symmetric spacetimes with weak regularity,
Class.\ Quantum Grav.\ 28 (2011), 145019--145035.

\bibitem{LR-2} 
{\sc H. Lindblad and I. Rodnianski,}
The global stability of Minkowski spacetime in harmonic gauge,
Ann. of Math. 171 (2010), 1401--1477.

\bibitem{MarsSenovilla}
\auth{ M. Mars and J.M. Senovilla,}
Geometry of general hypersurfaces in space- time: junction conditions, 
Class. Quantum Grav., 10 (1993), 1865--1897.

\bibitem{Penrose0}
\auth{R.~Penrose,}
A remarkable property of plane wave in general relativity,
Rev.\ Modern Phys.\ 37 (1965), 215--220.

\bibitem{Penrose}
\auth{R.~Penrose,}
The geometry of impulsive gravitational waves,
in: ``General Relativity, Papers in honour of J.L.~Synge'',
ed.\ L.~O'Raifeartaigh, 1972, Clarendon Press, Oxford, pp.~101--115.

\bibitem{PenroseCCC1}
\auth{R.~Penrose,}
Before the big bang: an outrageous new perspective and its implications for particle physics,
in: ``EPAC 2006 proceedings'',
ed.\ C.R.~Prior, 2006, European Physical Society Accelerator Group, Edinburgh, pp. 2759--2762.

\bibitem{Rendall:2008}
\auth{A.D.~Rendall,}
{\sl Partial differential equations in general relativity,}
Oxford Graduate texts in Math., Oxford Univ.\ Press, 2008.

\bibitem{RendallWeaver}
\auth{A.D.~Rendall and M.~Weaver,}
Manufacture of Gowdy spacetimes with spikes,
Class.\ Quantum Grav.\ 18 (2001), 2959--2975.

\bibitem{SteinhardtTurok2004}
\auth{P.J.~Steinhardt and N.~Turok,}
Beyond inflation: a cyclic universe scenario,
Phys.\ Scripta T 117 (2005), 76.

\bibitem{Tod}
\auth{K.P.~Tod,}
Isotropic cosmological singularities,
in ``The conformal structure of spacetime: Geometry, Analysis, Numerics'', Springer Verlag, 2002, pp.~123--134.

\bibitem{Wang} 
\auth{Q. Wang}, 
An intrinsic hyperboloid approach for Einstein Klein-Gordon equations, 
Preprint ArXiv: 1607.01466. 

\bibitem{Wilson-Ewing-LQC}
\auth{E.~Wilson-Ewing,}
The loop quantum cosmology bounce as a Kasner transition,
Class.\ Quant.\ Grav.\ 35 (2018), 065005.

\end{thebibliography}
\end{document}